\documentclass[preprint]{revtex4}

\usepackage{graphicx}%
\usepackage{dcolumn}
\usepackage{amsmath}

\makeatletter
\def\btt#1{\texttt{\@backslashchar#1}}%
\DeclareRobustCommand\bblash{\btt{\@backslashchar}}%
\makeatother

\begin{document}

\title{Cosmological Anisotropy and the Cycling of Universe}

\author{Jian-gang Hao}

\author{Xin-zhou Li}\email{kychz@shtu.edu.cn}

\affiliation{ Shanghai United Center for Astrophysics, Shanghai
Normal University, 100 Guilin Road, Shanghai 200234 , China\\
School of Science, East China University of Science and
Technology, 130 Meilong Road, Shanghai 200237, China
}%

\date{\today}

\begin{abstract}
Abstract : In this paper, we analyze the anisotropy of the scale
factor in the Kantowski-Sachs spacetime. We show that the
anisotropy will not increase when the expansion rate is greater
than certain values while it will increase when the expansion rate
is less than that value or the Universe is contracting. It is
manifested that the matter dominated and radiation dominated era
favor the flat spacetime if the anisotropy does not develop
significantly. The relation between the cosmological anisotropy
and the red-shift of the supernovae, which could be used to verify
the anisotropy through the observation, has been derived.
\end{abstract}

\pacs{98.80.-k, 04.20.-q}

\maketitle

\vspace{0.4cm} \noindent\textbf{1. Introduction} \vspace{0.4cm}

Recent results from WMAP satellite offer a lot of precise
information about the Universe\cite{wmap}, while not yet resolve
many problems, among which is that they can not be used to
discriminate between cosmological models such as cyclic model and
conventional Big Bang plus inflation model\cite{stscience}. The
cyclic model for the Universe has been proposed by Steinhardt and
Turok\cite{steinhardt1}, which is based on the spin-off of
ekpyrotic scenario\cite{steinhardt2}whose starting point is the
unified theory known as heterotic M-theory\cite{witten}. Comparing
with the consensus model of Universe, the big bang was replaced
with a phase transition from contraction phase to expansion phase
in the new picture. The expansion period of the cycle contains a
period of radiation and matter domination ensued by an extended
period of cosmic acceleration at low energy, which establishes the
flat and vacuous initial condition required for the ekpyrosis and
for removing the entropy, black holes and other debris produced in
the preceding cycle. One of the remarkable feature of cyclic model
is that it avoids the unnatural magnitude difference (100 orders)
of the two period of accelerating expansion, inflation and current
observed cosmic acceleration. Also, it provides a new mechanism
for the generation of a scale-invariant spectrum of density
perturbation and an explanation for the big bang.

In this paper, we analyze the development of cosmological
anisotropy during the expansion and contraction phase in
Kantowski-Sachs Universe. Different from conventional study in
this field\cite{kantowski}, we try a new approach to investigate
the evolution of anisotropy and find that in a typical power-law
expansion, the anisotropy will increase rapidly if the expansion
rate $a(t)$ is slower than $a(t)\sim t^{\frac{1}{3}}$ for the flat
Universe, or $a(t)\sim t$ for open and closed Universe, or the
Universe is undergoing an contraction. If $a(t)$ expands/contracts
in an exponential manner, the anisotropy will decrease/increase
exponentially. The radiation dominated and matter dominated era,
which correspond to $a(t)\sim t^{\frac{1}{2}}$ and $a(t)\sim
t^{\frac{2}{3}}$, will not produce significant anisotropy when and
only when the Universe is spatially flat, which is in agreement
with the predication of inflation theory\cite{inflation} as well
as the very recent WMAP observations\cite{stscience}.

\vspace{0.4cm} \noindent\textbf{2. The Evolution of Anisotropy}
 \vspace{0.4cm}

The anisotropic Robertson-Walker Universe has been widely studied
\cite{kantowski} and the corresponding metric can be written as

\begin{equation}\label{metric}
ds^2=-dt^2+b^2(t)dr^2+a^2(t)(d\theta^2+S(\theta)^2d\phi^2)
\end{equation}
\noindent where

\begin{equation}\label{s}
 S(\theta)=
  \begin{cases}
    \sin\theta & for \hspace{1cm} k=1, \\
    \theta & for \hspace{1cm} k=0,\\
    \sinh\theta & for \hspace{1cm} k=-1
  \end{cases}
\end{equation}

\noindent and $k=1, 0, -1$ corresponds to the closed, flat and
open universe. Conventional discussions on the above type of
spacetime focus on the quantities know as usual expansion
$\Theta$, shear $\sigma$ and 3-curvature $^{(3)}R$, which are
defined as

\begin{equation}\label{definequantity}
\Theta=\frac{\dot{b}}{b}+\frac{2\dot{a}}{a}, \hspace{1cm}
\sigma=\frac{1}{\sqrt{3}}(\frac{\dot{b}}{b}-\frac{\dot{a}}{a}),
\hspace{1cm} ^{(3)}R=\frac{2k}{a^2}
\end{equation}

Those with zero and negative curvature are just axisymmetric
Bianchi type I and III Universe while the positive curvature
model, or the closed the anisotropic Universe model, is referred
to as the Kantowski-Sachs Universe. Although only the closed
models fall outside of the Bianchi classification, yet one can
refer to them all as Kantowski-Sachs models for
convenient\cite{barrow}. In this paper, we try a new approach to
this models and obtain some interesting results. The Einstein
equations correspond to the above setup are:
\begin{equation}\label{einstein1}
(\frac{\dot{a}}{a})^2+2\frac{\dot{a}}{a}\frac{\dot{b}}{b}+\frac{k}{a^2}=\kappa\rho
\end{equation}

\begin{equation}\label{einstein2}
 2\frac{\ddot{a}}{a}+(\frac{\dot{a}}{a})^2+\frac{k}{a^2}=-\kappa p
\end{equation}

\begin{equation}\label{einstein3}
  \frac{\ddot{a}}{a}+\frac{\ddot{b}}{b}+\frac{\dot{a}}{a}\frac{\dot{b}}{b}=-\kappa
  p
\end{equation}

\noindent where $\kappa=8\pi G$, $p$ and $\rho$ are the pressure
and energy density of the perfect fluid respectively. The energy
conservation of the perfect fluid is expressed as
\begin{equation}\label{conservation}
\frac{d \rho}{dt}=-\Theta (\rho+p)
\end{equation}

\noindent Substitute Eq.(\ref{einstein3})into
Eq.(\ref{einstein2}), we have
\begin{equation}\label{transform}
\frac{\ddot{a}}{a}-\frac{\ddot{b}}{b}+\frac{\dot{a}}{a}\Bigg(\frac{\dot{a}}
{a}-\frac{\dot{b}}{b}\Bigg)+\frac{k}{a^2}=0
\end{equation}

\noindent To study the evolution of the anisotropy, it is
convenient to introduce a new quantity $\delta$ that is defined as
$\delta\equiv\frac{b-a}{a}$. Then, the metric tensor(\ref{metric})
can be written as
\begin{equation}\label{newmetric}
ds^2=-dt^2+a^2(t)(1+\delta)^2dr^2+a^2(t)(d\theta^2+\theta^2d\phi^2)
\end{equation}

\noindent where $\delta$ can also be interpreted as a perturbation
of the scale factor if it is small. Now, we can rewrite
Eq.(\ref{transform}) in terms of $\delta$ as

\begin{equation}\label{finaleuqation}
\ddot{\delta}+3\frac{\dot{a}}{a}\dot{\delta}-\frac{k}{a^2}=0
\end{equation}

\noindent It is not difficult to find the first integral of
Eq.(\ref{finaleuqation}) as
\begin{equation}\label{firstintegral}
\frac{\dot{\delta}(t)}{\dot{\delta}(t_0)}=\frac{1}{\dot{\delta}(t_0)a^3(t)}
\int_{t_0}^{t}k a(t) dt + \bigg[\frac{a(t_0)}{a(t)}\bigg]^3
\end{equation}

\noindent where $\dot{\delta}(t_0)$ is the initial "speed" of the
anisotropy and $a(t_0)$ is the initial value of the scale factor
at $t=t_0$. In the following, we investigate the generic evolution
of the anisotropy in power law expansion and exponential
expansion. For the power law expansion $a(t)=a(t_0)t^q$, the
solutions for Eq.(\ref{firstintegral}) are:

\noindent (i) If $q\neq\pm1$ and $q\neq\frac{1}{3}$

\begin{eqnarray}\label{deltageneral1}
\delta(t)=&&\delta(t_0)-\frac{k
t_0^{2}}{(q+1)(2-2q)a^2(t_0)}-\left[\frac{\dot{\delta}(t_0)t_0}{1-3q}-
\frac{k t_0^{2}}{(q+1)(1-3q)a^2(t_0)}\right]\\\nonumber
&&+\left[\dot{\delta}(t_0)t_0^{3q}-
\frac{kt_0^{3q+1}}{(q+1)a^2(t_0)}\right]\frac{t^{1-3q}}{1-3q}+\frac{kt_0^{2q}t^{2-2q}}{(q+1)(2-2q)a^2(t_0)}
\end{eqnarray}

\noindent (ii) If $q=-1$
\begin{equation}\label{deltageneral2}
\delta(t)=\delta(t_0)+\frac{kt_0^2}{16a^2(t_0)}-\frac{\dot{\delta}(t_0)t_0}{4}+
\frac{kt^4}{16a^2(t_0)t_0^2}(4\ln\frac{t}{t_0}-1)+\frac{\dot{\delta}(t_0)t^4}{4t_0^3}
\end{equation}

\noindent (iii) If $q=1$
\begin{equation}\label{deltageneral3}
\delta(t)=\delta(t_0)+\frac{1}{2}[\dot{\delta}(t_0)t_0-\frac{kt_0^2}{2a^2(t_0)}]
+\frac{kt_0^2}{2a^2(t_0)}\ln(\frac{t}{t_0})-\frac{1}{2}[\dot{\delta}(t_0)t_0^3-\frac{kt_0^4}{2a^2(t_0)}]t^{-2}
\end{equation}

\noindent (iv) If $q=\frac{1}{3}$
\begin{equation}\label{deltageneral4}
\delta(t)=\delta(t_0)-\frac{9kt_0^{2}}{16a^2(t_0)}+\frac{9kt_0^{2/3}}{16a^2(t_0)}t^{4/3}+[\dot{\delta}(t_0)t_0-\frac{3kt_0^{2}}{4a^2(t_0)}]\ln
\frac{t}{t_0}
\end{equation}

\noindent Now, let's briefly discuss the above solutions.

\noindent In case (i), one can easily find that when $k=0$, the
anisotropy $\delta$ will increase with time if $q<\frac{1}{3}$
while it will decrease if $q>\frac{1}{3}$. However, when $k=\pm1$,
the anisotropy $\delta$ will increase with time if $q<1$ while it
will decrease if $q>1$. Clearly, the evolution of $\delta$ is
different for closed Universe ($k=1$) and open Universe ($k=-1$),
and the difference is dependent on the choice of the initial
$a(t_0)$, $t_0$, $k$ and $\dot{\delta}(t_0)$.

\noindent In case (ii), one can find that $\delta$ will always
increase with time but obviously in different manners, which are
also dependent on the choice of the initial $a(t_0)$, $t_0$, $k$
and $\dot{\delta}(t_0)$.

\noindent In case (iii), one can find that when $k=0$, $\delta$
will always decrease with time. But when $k=\pm1$, $\delta$ will
always increase in different manners dependent on $a(t_0)$, $t_0$,
$k$ and $\dot{\delta}(t_0)$.

\noindent In case (iv), one can find that $\delta$ will always
increase with time but in different manners for different
$a(t_0)$, $t_0$, $k$ and $\dot{\delta}(t_0)$.

It is also interesting to consider the anisotropy in the
exponential expansion $a(t)=a(t_0)\exp[H(t-t_0)]$. It is not
difficult to find that the $\delta$ evolves as
\begin{equation}\label{deltinexp}
\delta(t)=\delta(t_0)+\frac{k}{6H^2a^2(t_0)}+\frac{\dot{\delta}(t_0)}{3H}
-\frac{k}{2H^2a^2(t_0)}e^{-2H(t-t_0)}+\left[\frac{k}{3H^2a^2(t_0)}-
\frac{\dot{\delta}(t_0)}{3H}\right]e^{-3H(t-t_0)}
\end{equation}

It is quite clear that $\delta(t)$ will decrease with $t$ if
$a(t)$ expand exponentially, and it will increase if $a(t)$
contracts exponentially. So far, we can conclude that for the
standard Big Bang cosmology, the cosmological anisotropy will
increase during the radiation dominated and matter dominated era
if the Universe is open or close. It can only survive for the
spatially flat cosmology, which is an important prediction of
inflation and confirmed by the recent WMAP data. In the subsequent
section, we will devote our discussions mainly to the case in a
spatially flat spacetime.

\vspace{0.4cm} \noindent\textbf{3. the Evolution of Anisotropy in
Flat Spacetime and Cyclic Model }
 \vspace{0.4cm}

From Eq.(\ref{firstintegral}), by setting $k=0$, we have the first
integral of Eq.(\ref{finaleuqation}) in flat Universe as

\begin{equation}\label{firstinterflat}
\frac{\dot{\delta}(t)}{\dot{\delta}(t_0)}=
\bigg[\frac{a(t_0)}{a(t)}\bigg]^3
\end{equation}

\noindent From Eq.(\ref{firstinterflat}), we have
$\frac{\dot{\delta}(t)}{\dot{\delta}(t_0)}>0$, therefore
$\delta(t)$ is a monotonic function. When $\dot{\delta}(t_0)>0$,
$\delta(t)$ increases monotonically; when $\dot{\delta}(t_0)<0$,
$\delta(t)$ decreases monotonically.

It does not lose generality that we assume $b(t_0)>a(t_0)$, i.e.,
$\delta(t_0)\geq 0$ for simplicity. The generalized scale factor
$a(t)$ is determined by the pressure and energy density of the
perfect fluid in the Einstein equations
(\ref{einstein1})-(\ref{einstein3}). For example, we can apply
these equations specifically to the initial stage of cosmological
evolution which  is assumed to be governed by the ordinary scalar
field or the rolling tachyon filed. In this paper, these equations
are applicable to matter with energy momentum tensor of arbitrary
form so that we should discuss the varied form of $a(t)$. From
Eq.(\ref{firstinterflat}), we have
\begin{equation}\label{delta}
 \delta(t)\leq\delta(t_0)+|\dot{\delta}(t_0)|\cdot |\int_{t_0}^{t}\Big[\frac{a(t_0)}
 {a(t)}\Big]^3dt|
\end{equation}

\noindent By using the consensus Cauchy criterion, we discuss the
convergence of definite integral in Eq.(\ref{delta}). At $t\gg
t_0$, we rewrite $a(t)$ as $\frac{f(t)}{t^q}$, if $q>\frac{1}{3}$
and $f(t)\leq constant \leq +\infty$, then the integral is
convergent; If $q\leq\frac{1}{3}$ and $f(t)>constant\geq 0$, then
the integral is divergent. This argument can be extended to the
following form: if $\bigg[\frac{a(t_0)}{a(t)}\bigg]^3=f(t)\cdot
g(t)$ for $t\gg t_0$, $\int_{t_0}^{\infty}g(t)dt$ is a convergent
integral and $f(t)$ is a monotonic and boundary function, then
$I\equiv \int\bigg[\frac{a(t_0)}{a(t)}\bigg]^3dt$ is a convergent
integral. The observed expansion of the universe and the observed
cosmic black body radiation provide the empirical basis for a
Friedmann-Lemaitre-Robertson-Walker model of the universe,
sometimes called the isotropic model. Therefore, we assume that
the initial values of anisotropy must be tiny as compared with the
scale factor, i.e., $\delta(t_0)\ll 1$ and
$|\frac{\dot{b}}{b}-\frac{\dot{a}}{a}|_{t=t_0}\ll 1$. Therefore,
the anisotropy degree $\delta(t)$ is still tiny at the late time
when $I$ is a convergent integral. We now have proved a lemma that
is stated as follows:

$Lemma$ 1. There is an anisotropic perturbation at $t=t_0$, which
is described by $\delta(t_0)$ and $\dot{\delta}(t_0)$ in the
isotropic universe with the scale factor $a(t)$. The anisotropy
degree $\delta(t)$ is still tiny at the late time evolution when
the expansion rate is greater than certain value while it will
become large when the expansion rate is equal or smaller than that
value or the universe is undergoing an contraction stage appeared
in the cyclic model.

Now, let's consider the power law expansion of the scale factor
$a(t)=a(t_0)\frac{t}{t_0}^{q}$. The cases $q=1/2$, $q=2/3$ and
$q>1$ represent the radiation dominated era, matter dominated era
and accelerating expansion era respectively. The $q<0$ case
corresponds to the contraction phase in the cyclic universe model.
Using $a(t)=a(t_0)\frac{t}{t_0}^{q}$, one can easily obtain
\begin{equation}\label{deltaresult}
\delta(t)=
\begin{cases}
\delta(t_0)-\frac{\dot{\delta}(t_0)t_0}{1-3q}+\frac{\dot{\delta}(t_0)t_0^{3q}}{1-3q}t^{1-3q},
\hspace{1.4cm} q\neq \frac{1}{3} \\
\delta(t_0)+\dot{\delta}(t_0)t_0\ln(\frac{t}{t_0}), \hspace{2.6cm}
q=\frac{1}{3}
\end{cases}
\end{equation}

\noindent If the condition $q>1/3$ is satisfied, we have
\begin{equation}\label{condition}
\delta(t_0)-\frac{|\dot{\delta(t_0)}|}{3q-1}t_0\leq \delta(t)\leq
\delta(t_0)+\frac{|\dot{\delta(t_0)}|}{3q-1}t_0
\end{equation}
\noindent so that the anisotropy degree $\delta(t)$ is small
forever, and is specified by its initial value. In the radiation
and matter dominated era as well as the accelerating expansion
era, the expansion satisfy the condition $q>1/3$. Therefore, the
evolution of anisotropy degree has been so significantly damped
that it could not be detected by current observation. When the
expansion is slower than $t^{1/3}$, or the universe is
contracting, the anisotropy degree will increase rapidly with
time.

\noindent When the scale factor is in the exponential expansion or
contraction $a(t)=a(t_0)e^{\pm H(t-t_0)}$, we will have
\begin{equation}\label{exp}
\delta(t)=\delta(t_0)\pm\frac{\dot{\delta}(t_0)}{3H}[1-e^{\pm3H(t_0-t)}]
\end{equation}

\noindent where $H$ is the Hubble constant and the plus and minus
signs represent the expansion and contraction respectively. It is
obvious that the anisotropy degree $\delta(t)$ satisfies

\begin{equation}\label{condition1}
\delta(t_0)-\frac{|\dot{\delta(t_0)}|}{3H}\leq \delta(t)\leq
\delta(t_0)+\frac{|\dot{\delta(t_0)}|}{3H}
\end{equation}

\noindent when the scale factor is expanding exponentially, while
the anisotropy degree will increase rapidly with time when the
scale factor contracts exponentially.

In cyclic model, the cyclic scenario can be effectively described
in terms of the evolution of a scalar field in a specific
potential $V(\phi)$. The essential difference is in the form of
the potential and the couplings between the scalar field, matter
and radiation\cite{steinhardt1}. In the anisotropic R-W
metric(\ref{newmetric}), the action for the model can be expressed
as
\begin{equation}\label{action}
S=\int d^4x\sqrt{-g}\left[\frac{1}{2\kappa}{\cal
R}-\frac{1}{2}(\partial\phi)^2-
V(\phi))+\beta^4(\phi)(\rho_M+\rho_R)\right]
\end{equation}

\noindent where $g$ is the determinant of the metric $g_{\mu \nu}$
and ${\cal R}$ is the Ricci scalar. The coupling $\beta(\phi)$
between $\phi$ and the matter ($\rho_M$) and radiation ($\rho_R$)
densities is crucial because it allows the densities to remain
finite at the big crunch/big bang transition. Varying the
action(\ref{action}), one can obtain the Einstein equation up to
the linear order of the anisotropic perturbation as
\begin{equation}\label{einstein11}
3H^2+2H\dot{\delta}=\kappa \left( \frac{1}{2} \dot{\phi}^2 +V +
\beta^4 \rho_R + \beta^4 \rho_M \right)
\end{equation}

\begin{equation}\label{einstein22}
 2\frac{\ddot{a}}{a}+H^2=- \kappa \left(\dot{\phi}^2 -V +
\beta^4 \rho_R +{1\over 2} \beta^4 \rho_M \right)
\end{equation}

\begin{equation}\label{einstein33}
  2\frac{\ddot{a}}{a}+3H\dot{\delta}+H^2+\ddot{\delta}=- \kappa \left(\dot{\phi}^2 -V +
\beta^4 \rho_R +{1\over 2} \beta^4 \rho_M \right)
\end{equation}

\noindent The fluid equation of motion for matter or fluid is
\begin{equation}\label{flueq}
\frac{d \rho_i}{d
t}=\left[-3\frac{\dot{\hat{a}}}{\hat{a}}-\dot{\delta}\right](\rho_i+p_i)
\end{equation}

\noindent where $\hat{a}=\beta(\phi) a$ and $\beta(\phi)$ is such
chosen that the quantity $\hat{a}$ is finite when $a$ approaches
zero.

Now, we can consider the cosmological anisotropy in a similar
fashion as that in the previous section of this paper. It is clear
that the evolution equation for the anisotropy can also be
expressed as Eqs.(\ref{deltaresult}) and (\ref{exp}) for the power
law and exponential expansion/contraction respectively. Therefore
the conclusions in the former part of this section are still held
true. When the universe is in the contraction phase, the
anisotropy will develop to be very significant. It must be pointed
out that the increase of anisotropy does not necessarily exclude
the cycling of Universe as described in the cyclic model if the
initial "speed" of anisotropy $\dot{\delta}(t_0)$ is extremely
small so that the anisotropy will not develop to be very
significant even after a long period increase as shown in
Eq.(\ref{deltaresult}). This is physically possible because the
lasting accelerated expansion before the contraction phase will
reduce the initial "speed" of anisotropy. On the other hand, one
can also consider a different coupling between the scalar field
and the matter as well as the radiation, for example, by taking
into account of the higher order anisotropic effects and the
corresponding coupling that dependent on direction, to avoid the
the increase of anisotropy. The later possibility will be
attempted in a preparing work.

In the following section, we will correlate the anisotropy with
the red-shift of the supernovae, which might be a way to identify
the anisotropy not only when it increases, but also when it damps.

\vspace{0.4cm} \noindent\textbf{4. Relate the Anisotropy with the
Red-shift}
 \vspace{0.4cm}

 The red-shift data of supernovae has been used to reconstruct the
potential\cite{starobinski, turner1, turner2}or the equation of
state\cite{hao} of the field that drive the expansion of the
universe. Now, we generalize the red-shift relations in isotropic
flat R-W spacetime by introducing the different red-shift in
different directions. In our set-up here, we consider the simple
case that the red-shift is different in the $x$ direction and the
$y,z$ plane. Accordingly, we introduce the following definitions:
\begin{equation}\label{xred}
 x(Z_x)=\int_{t(Z_x)}^0\frac{du}{b(u)}=\int_0^{Z_x}\frac{d\xi}{H_x(\xi)}
\end{equation}

\begin{equation}\label{yred}
 y(Z_y)=\int_{t(Z_y)}^0\frac{du}{a(u)}=\int_0^{Z_y}\frac{d\xi}{H_y(\xi)}
\end{equation}

\noindent where $x(Z_x)$ and $y(Z_y)$ are the coordinate distance
in $x$ and $y$ directions respectively. $Z_x$ and $Z_y$ are the
red-shift in $x$ and $y$ direction respectively. $H_x(Z)$ and
$H_y(Z)$ are defined as

\begin{equation}\label{difx}
(\frac{\dot{b}}{b})^2=H_x(Z_x)^2=\frac{1}{(dx/dZ_x)^2}
\end{equation}

\begin{equation}\label{dify}
(\frac{\dot{a}}{a})^2=H_y(Z_y)^2=\frac{1}{(dy/dZ_y)^2}
\end{equation}
$H_x(Z)\equiv\frac{\dot{b}}{b}$ and
$H_y(Z)\equiv\frac{\dot{a}}{a}$. Since the $z$ direction has the
same scale factor as that in $y$ direction, we consider the
corresponding coordinate distance in $z$ direction is the same as
in $y$ direction. Obviously, the above definitions is a direct
generalization of the red-shift in the isotropic spacetime

\begin{equation}\label{rred}
r(Z)=\int^{t_0}_{t(Z)}\frac{du}{a(u)}=\int^Z_0\frac{d\xi}{H(\xi)}
\end{equation}

\noindent It is not difficult to prove that the relations between
$t$ and $Z_x$ and $Z_y$ are
\begin{equation}\label{relationx}
\frac{dZ_x}{dt}=-(1+Z_x)H_x(Z_x)=-(1+Z_x)\frac{dx}{dZ_x}
\end{equation}

\begin{equation}\label{relationy}
\frac{dZ_y}{dt}=-(1+Z_y)H_y(Z_y)=-(1+Z_y)\frac{dy}{dZ_y}
\end{equation}

To correlate the red-shift $Z_x$ and $Z_y$ with the anisotropy
$\delta$, we can consider Eqs.(\ref{firstintegral}) and (
\ref{difx}). From Eq.(\ref{dify}), one have
\begin{equation}\label{intere}
d\ln a=\frac{dt}{dy/dZ_y}
\end{equation}

\noindent Together with Eq.(\ref{firstintegral}), we have

\begin{eqnarray}\label{deltaredshift}
\dot{\delta}(Z_x, Z_y)&&\equiv f(Z_x, Z_y)\\\nonumber
&&=\dot{\delta_0}\exp[\int\frac{3(dx/dZ_x)}{(dy/dZ_y)(1+Z_x)}dZ_x+3\int\frac{dZ_y}{(1+Z_y)}]
\end{eqnarray}

\noindent Therefore, the correlation equations could be expressed
as:

\begin{equation}\label{corr1}
\frac{\partial \delta}{\partial Z_x}+\frac{\partial
\delta}{\partial
Z_y}\frac{(1+Z_y)(dy/dZ_y)}{(1+Z_x)(dx/dZ_x)}=-\frac{f(Z_x,
Z_y)}{(1+Z_x)(dx/dZ_x)}
\end{equation}

\begin{equation}\label{corr2}
\delta(Z_x, Z_y)=-\int[\frac{f(Z_x,
Z_y)dZ_x}{(1+Z_x)(dx/dZ_x)}+\frac{f(Z_x,
Z_y)dZ_y}{(1+Z_y)(dy/dZ_y)}]
\end{equation}

It is obvious that once we know the coordinate distance $x$ and
$y$ as the function of red-shift $Z_x$ and $Z_y$, we will be able
to obtain the anisotropy $\delta$ through the correlation
relations Eqs.(\ref{corr1}) and (\ref{corr2}).

\vspace{0.4cm} \noindent\textbf{4. Discussion and Conclusion}
 \vspace{0.4cm}

In this paper, we study the evolution of cosmological anisotropy
in Kantowski-Sachs spacetime. We show that the anisotropy will
increase significantly if the expansion rate is slower than a
critical value, which is different for different type
Universe(Closed, open or flat). We also show that the expansion in
the conventional big bang cosmology, the radiation dominated era,
matter dominated era and the current accelerating expansion era,
is fast enough to damp the anisotropy if the Universe is spatially
flat, which is in agreement with the theoretical predication of
inflation theory as well as current observations. But if the
Universe is not flat, the anisotropy will develop to be
significant during the radiation and matter dominated era. This
itself could be considered as a important theoretical support to
the inflation theory. In the cyclic Universe model, the anisotropy
will increase when the Universe expands slower than the critical
value or contracts. But, as we explained in section 3, this could
be alleviated by the tuning of the condition before Universe
expands slower than the critical value, or by considering
different and direction-dependent coupling between the scalar
field and radiation as well as matter. Especially, it needs to
point out that in the cyclic model, $q$ does not remain small all
the way to the bounce. Rather, $q$ is greater than $\frac{1}{3}$
once the branes get to within a few Planck
lengths\cite{steinhardt3}. So, the above increase of anisotropy
could also be considered as a mechanism employed by the cyclic
model to seed acceptable anisotropy observed in the present
Universe.

On the other hand, observation of the red-shift of SNeIa has been
proved to be an important way to probe the
Universe\cite{Perlmutter}. We derive the relation between the
cosmological anisotropy and the red-shift, which indicates that
one can gain some information about the anisotropy by fitting the
red-shift data. When the Universe slows down or contracts, we may
reconstruct the anisotropy through the red-shift data.

\vspace{0.8cm} \noindent ACKNOWLEDGMENTS

The authors thank Professor P Steinhardt and J Barrow for helpful
comments. This work was partially supported by National Nature
Science Foundation of China under Grant No. 19875016 and
Foundation of Shanghai Development for Science and Technology
under Grant No. JC 14035.

\end{document}